\documentclass[screen,sigconf]{acmart}
\usepackage{color}
\usepackage{booktabs} 
\usepackage{url}
\usepackage{graphicx}
\usepackage{subfigure}
\usepackage{multirow}
\usepackage{xspace}
\usepackage{enumitem} 
\usepackage[normalem]{ulem}
\usepackage{soul}
\usepackage{balance}
\usepackage{comment}
\usepackage{dirtytalk}
\usepackage{makecell}

\usepackage{amsmath}

\usepackage{amssymb}

\usepackage{amsfonts}

\usepackage{algorithmic}
\usepackage{textcomp}
\usepackage{xcolor}
\usepackage{tcolorbox}

\newcommand{\chen}[1]{\textcolor{red}{#1}}

\begin{document}

\title{Software Engineers’ Response to Public Crisis: Lessons Learnt from Spontaneously Building an Informative COVID-19 Dashboard}

\author{Han Wang}
\email{han.wang@monash.edu}
\affiliation{%
  \institution{Faculty   of   Information   Technology,   Monash   University}
  \country{Australia}
}
\author{ Chao Wu}
\email{benyunlongwu@gmail.com}
\affiliation{%
  \institution{Faculty   of   Information   Technology,   Monash   University}
  \country{Australia}
}
\author{Chunyang Chen}
\email{ chunyang.chen@monash.edu}
\affiliation{%
  \institution{Faculty   of   Information   Technology,   Monash   University}
  \country{Australia}
}
\authornote{Corresponding authors}

\author{Burak Turhan}
\email{burak.turhan@oulu.fi}
\affiliation{%
  \institution{Faculty of Info. Tech. and Electrical Eng., University of Oulu}
  \country{Finland}
}

\author{Shiping Chen}
\email{shiping.chen@data61.csiro.au}
\affiliation{%
  \institution{Data61 CSIRO}
  \country{Australia}
}
\author{Jon Whittle}
\email{jon.whittle@data61.csiro.au}
\affiliation{%
  \institution{Data61 CSIRO}
  \country{Australia}
}

\copyrightyear{2022}
\acmYear{2022}
\setcopyright{acmlicensed}\acmConference[ICSE-SEIS'22]{Software Engineering in Society}{May 21--29, 2022}{Pittsburgh, PA, USA}
\acmBooktitle{Software Engineering in Society (ICSE-SEIS'22), May 21--29, 2022, Pittsburgh, PA, USA}
\acmPrice{15.00}
\acmDOI{10.1145/3510458.3513010}
\acmISBN{978-1-4503-9227-3/22/05}

\begin{abstract}
The Coronavirus disease 2019 (COVID-19) outbreak quickly spread around the world, resulting in over 240 million infections and 4 million deaths by Oct 2021. 
While the virus is spreading from person to person silently, fear has also been spreading around the globe. 
The COVID-19 information from the Australian Government is convincing but not timely or detailed, and there is much information on social networks with both facts and rumors.
As software engineers, we have spontaneously and rapidly constructed a COVID-19 information dashboard aggregating reliable information semi-automatically checked from different sources for providing one-stop information sharing site about the latest status in Australia.
Inspired by the John Hopkins University COVID-19 Map, our dashboard contains the case statistics, case distribution, government policy, latest news, with interactive visualization.
In this paper, we present a participant's in-person observations in which the authors acted as founders of \url{https://covid-19-au.com/} serving more than 830K users with 14M page views since March 2020. 
According to our first-hand experience, we summarize 9 lessons for developers, researchers and instructors. These lessons may inspire the development, research and teaching in software engineer aspects for coping with similar public crises in the future. 

\end{abstract}


\maketitle
\section*{Lay Abstract}
The 2019 Coronavirus Disease (COVID-19) outbreak has spread rapidly around the world. By October 2021, it has caused more than 240 million infections and 4 million deaths. Although the world is acting against the virus, some information on the Internet has not been updated in time, and there are also many rumors on social media. Therefore, software engineers have developed COVID-19 information dashboards such as the Johns Hopkins University COVID-19 Map and the World Health Organization COVID-19 website to provide the public with one-stop reliable COVID-19 related information. The author has also developed a COVID-19 dashboard \url{https://covid-19-au.com/} that provides case statistics, case distribution, government policies, latest news, and interactive visualization during the pandemic in Australia. It has been popular since March 2020 and has provided 14 million page views to more than 830,000 users. In this paper, the authors discussed how they built the COVID-19 dashboard website and how they formed and managed a team of volunteers to help and maintain the project. More importantly, the authors summarized 9 lessons for developers, researchers and instructors based on experience. These courses may inspire them in development, research and teaching to deal with similar public crises in the future, and ultimately bring accurate information to public users more effectively.

\section{Introduction}
\label{sec:introduction}
Coronavirus disease 2019 (COVID-19)~\cite{zhou2020pneumonia} is an infectious disease caused by severe acute respiratory syndrome coronavirus 2 (SARS-CoV-2).
Since its first identification in December 2019, COVID-19 has spread globally, resulting in an ongoing pandemic. By the time we write this paper, there have been over 240 million confirmed cases and caused over 4 million lives lost in more than 192 countries and territories by Oct 2021. 
To avoid the transmission of the virus, over 90 governments have taken actions such as lockdowns, border closures, working from home arrangements, enforcing social distancing, which influenced over half of the world's population~\cite{web:euroNews}. The pandemic has a severe influence on society, which has also caused global economic disruptions, education interruptions, and discrimination. 

Under this unprecedented situation, it is usual for people to feel the urge to constantly check for the latest updates and information via government websites, news outlets, and social media. 
Considering that COVID-19 transmission, it is important to inform people frequently and accurately with the latest status of COVID-19, particularly in their locality. 
Furthermore, the information can help the public take proper actions, and also guide governments' policy making for maintaining the community health. 
Therefore, there's a need for a timely and trustworthy platform to provide stats and information on the COVID-19 virus to the public.

In Australia, the federal and state governments' public media releases every day since the beginning of the outbreak, including data of new case numbers and information on symptoms and prevention from the COVID-19 virus.
However, there are some limitations with the government updates: First, Australian states report their status separately without national criteria, making the data types inconsistent for each other. There are also some conflicts between the state government data and federal government data from time to time. Second, suitable interactive data visualization is needed to display the information, rather than the current static, plain text. Third, updates are not frequent enough, i.e., once a day and at different times in each state, and many details are missing. 

As an alternative, social media updates COVID-19 related information more frequently with fine-grained detail, but these channels also include numerous rumours and misinformation and can sometimes spread even faster than the truth~\cite{Tasnim2020,cinelli2020}. Also, it's hard to see a historical trend due to its high frequent update and limited content size.

\begin{figure*}
	\centering
	\includegraphics[width=0.95\textwidth]{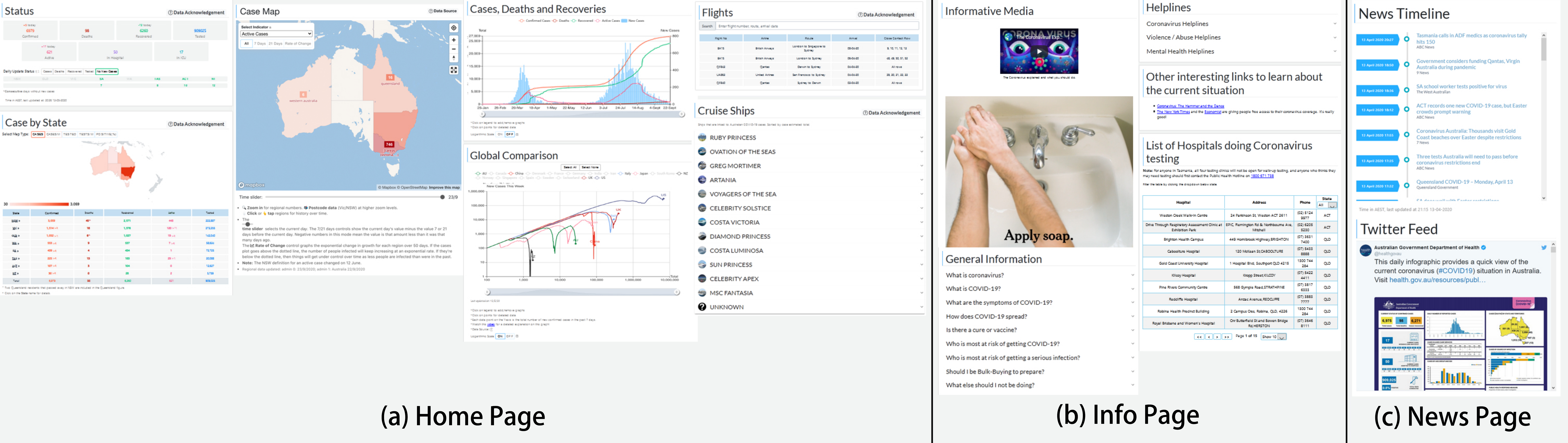}
	\caption{Different functionalities of our dashboard}
	\label{fig:overview}
	\vspace{-4mm}
\end{figure*}

To bridge the gap and respond to this crisis, we developed a COVID-19 information dashboard\footnote{\url{https://covid-19-au.com}} aggregating the information from different sources with intuitive data visualizations and frequent updates for providing people with the latest and the most accurate information. Our dashboard includes the latest COVID-19 status in each state (e.g., numbers of new cases, recovery, deaths, and trend), flight search, infection map, symptoms and prevention information, and latest news and tweets (as seen in Fig~\ref{fig:overview}) to address people's information needs.
To make the information more understandable, we apply different interactive data visualization techniques for different kinds of data with tables, line/bar charts, and also the advanced interactive map.

There are some similar dashboards around the world (such as COVID-19 global map from JHU~\cite{web:JHU}, 1Point3Acres COVID-19 tracking resource for North America~\cite{web:1Point3Acres}, DX Doctor's COVID-19 real-time report in China~\cite{web:DXY}), but our dashboard is the first one specifically targeting people in Australia with over 837K individual users and 14M page views. 
Through spontaneously and rapidly building this dashboard, we have accumulated much first-hand experience, including the requirement elicitation, user activity analysis, collaboratively coding and testing, and continuously releasing.



Based on the challenges and the experience, we summarise 9 lessons that would be helpful for the developers, researchers, and instructors respectively.
First, we discuss the usage of sensitive data, professional wording,  calming user feelings, and task allocation as lessons for developers when developing the COVID-19 dashboard. Second, for the academic researchers, we talk about the lessons related to how they can elicit people's information needs, analyse users' interactions on different devices, and understand users' interests in map visualizations by analysing the usage data and user feedback we collected. Third, we discuss how students can benefit from this project and how the instructors need to support the students from the perspective of software engineer instructors.
The lessons aim to help software engineers better understand how to deal with the pandemic as different roles.
In particular, they may guide software developers in developing similar applications for future crisis events, benefit researchers in conducting studies in this area, and inspire instructors in practical teaching about software engineering.

The COVID-19 dashboard is a great example of the importance of deeply thinking through what we call human values in software design. Human values - such as inclusion, diversity, social responsibility - are things that we as a society value, but they have often been paid little attention in software engineering~\cite{perera2020study}. In the case of a dashboard presenting important data on an emerging pandemic, it is critical that the design of the system thinks through possible negative consequences - for example, presenting the information in one way versus another may influence citizens' behaviours.


The contribution of this work is summarized as follows:
\begin{itemize}
	\item This is a timely social-good project for developing a COVID-19 information dashboard within a limited time for solving information needs of the society during this global pandemic.
	\item We document the first-hand development experience of an information dashboard for COVID-19 with great social impact on millions of people in Australia.
	\item During the dashboard development, we distill important lessons for developers, researchers, and instructors based on our own experience, and these lessons may guide them in the software engineer area and deliver value to the public.
\end{itemize}

The rest of the paper is organized as follows. Section 2 provides the background of the efforts the developers have made to fight against the pandemic, and some dashboards that have been released for the COVID-19. We introduce the dashboard characteristics, the huge impact it has made to the public, and the overall steps of how the authors developed it in section 3. In section 4, we summarize a total of 9 lessons learned during the development of the COVID-19 dashboard. We conclude our work in section 5 and discuss the improvements that could be made to the dashboard.

\section{Background}
Software systems increasingly impact our society. This is especially true for AI systems, which can replace humans as decision makers, but it is also true for more mundane software systems. Essentially, any software system has the potential for impact (positive or negative) on society and there is an increasing recognition that developers need to understand these impacts as part of the software design process~\cite{whittle2019case}. During the COVID-19 pandemic, developers are mobilising around the coronavirus outbreak to find solutions to deal with the crisis using different software technologies.

\label{sec:background}

\subsection{Developers' Efforts in Fighting the Virus}

The Information Technology community is one of the important forces in fighting with COVID-19. Various innovations and solutions ranging from contact tracing applications to data platforms were developed in response to different aspects of the pandemic. Researchers have also reviewed the proposed applications and taxonomies~\cite{whitelaw2020applications, javaid2020industry, budd2020digital}. Based on our experience and the existing research, we categorised such applications into 5 types:

\textbf{Symptom Tracking} applications that analyse symptom data from suspected and confirmed patients. Governments of countries such as Australia~\cite{web:auschecker}, Singapore~\cite{web:sgchecker}, and the US~\cite{web:uschecker} have rolled out their Self Checkers applications to direct individuals to appropriate resources. Some governments of countries such as Iceland~\cite{web:icechecker} and the UK~\cite{web:uschecker} use mobile technologies to collect data on patient-reported symptoms and triangulate to reveal information like how the virus is spreading and people who might have a risk that was exposed to the virus.

\textbf{Crowd Detection} applications that reduce the risk of virus transmission by monitoring the crowd in public areas. The Californian government worked with mobile app company Foursquare to track if public areas such as beaches and parks, were getting too crowded. In Denver, the Tri-County Health Department utilised data from ad-tech startup to monitor counties where the population on average strays more than 330 feet from home~\cite{web:foursquareData}. 
In the UK, the University of Newcastle uses data generated from smart city technology such as pedestrian flow, car park occupancy, and traffic to measure social distancing during the lock-down~\cite{web:ukCrowdMonitoring}. 

\textbf{Contact Tracing} applications identify and alert individuals who are close contacts of the confirmed patients. After Singapore launched the first COVID-19 contact tracing app ``Trace Together"~\cite{web:contactsgp} in March 2020, over 47 countries~\cite{web:contactlist} and territories have developed their own Contact Tracing app. Concerns and debates have arisen around the effectiveness of the software~\cite{web:contacteffective}, privacy issues~\cite{web:contactprivacy}, and project management failures~\cite{web:contactpm}. To tackle such problems, Apple and Google co-developed a joint API for contact tracing applications~\cite{web:contactApple}.

\textbf{Quarantine Daily Life} applications that help users cope with their life during the pandemic. Taiwan government uses a system called eMask to imposes quotas on masks~\cite{web:taiwanmask} to prevent hoarding. 
To help the self-employed people, financial technology professionals have created Covid Credit~\cite{web:CCredit}, a cloud-based tool developed with open banking technology to help people prove their income losses to the government. 
Microsoft's videoconferencing tool Teams uses artificial intelligence to cut out of the users' live photos and place them into a fixed position within a setting to create a more connected feel for the attendees~\cite{web:microsoftTeams}. Some developers even created applications that mimic office sounds~\cite{web:soundofColleagues} for users who miss working onsite.

\textbf{Information Aggregation and Sharing} applications aggregate and share COVID-19 data in real-time. Many applications collect and share COVID-19 datasets, such as COVID-19 Data Hub~\cite{guidotti2020covid}, DOMO tracker~\cite{web:dailypulse}, Our World In Data~\cite{web:ourworldindata} and Tableau COVID-19 global data~\cite{web:tablue}. 
The datasets were further visualised, analysed, and broadcasted. During December 2019, an AI company had sent the warnings of the outbreak by souring news reports and airline ticketing data~\cite{mccall2020covid}. 
In Europe, free live map of border crossing times for trucks was created to help Europe's supply chains to evaluate expected delays~\cite{web:truckboarder} based on traffic data. 
Some visualization applications~\cite{web:hongkonggis,web:harvard} including our dashboard focus on presenting COVID case information to help to provide a foundation for public policy decisions~\cite{web:ArcGIS}. 

\subsection{Information Dashboard about COVID-19}

A dashboard is a visual display of the most important information needed to achieve one or more objectives. Information was consolidated and arranged on a single screen to be monitored at a glance~\cite{few2007dashboard}.
There is a long history of using dashboards in response to disasters such as pandemics~\cite{wells2009post}, Hurricanes Katrina and Rita in the US~\cite{stone2007data} and storms in Brazil~\cite{kitchin2015knowing}. Nowadays, dashboards are widely used to effectively communicate disaster information~\cite{web:dashdisa} such as earthquakes~\cite{web:dashEarthquake}, hurricane~\cite{web:dashHurricane}, and bushfires~\cite{web:dashBushfire} to the public. 

Since the COVID-19 outbreak, the information dashboard has also been adopted for displaying the COVID-19 related information. 
For example, Johns Hopkins University developed a global COVID-19 dashboard~\cite{dong2020interactive} to display the reported cases on a daily time scale around the world. A number of similar dashboards are also developed by different entities and countries globally~\cite{wissel2020interactive, fernandez2020shiny, marivate2020use, roddiger2020responsible, web:WHO, web:DXY, web:1Point3Acres, berry2020open}. Other than the case number, dashboards were found to be used to display a variety of COVID data, including a clinical trial~\cite{thorlund2020real}, preventive measures~\cite{wimba2020dashboard}, social media sentiment level~\cite{pellert2020dashboard}, county vulnerability index~\cite{marvel2021covid}, and projection~\cite{florez2020online, mlocek2020forecasting}. Academic research in various disciplines has been generated based on these publicly available data dashboards~\cite{barone2020building, arora2020serotracker,koh2021pandemic, oronce2020association, parolini2020mathematical,li2021empirical}. 

However, along with the rapid emergence of various dashboards, some concerns have been raised. One research proposed the idea of "dashboard pandemic", stating that the dashboards are a "biopolitical technology of anxiety" and prompts national states of emergency, which lacks nuanced spatial, temporal, social, and epidemiological information. Thus not adequately addressing the uneven and unjust geographies of the present~\cite{everts2020dashboard}. An article addressed geopolitical considerations in maintaining, calling developers accountable to readers~\cite{muhareb2020tracking}. Efforts to address these issues have been made. For example, the US county-level dataset~\cite{killeen2020county} covers 300 variables from demographics to climate to make the dashboards more representative.

The authors developed an information dashboard\footnote{\url{https://covid-19-au.com}} for displaying the latest status of COVID-19 in Australia in early March 2020. Besides the basic data statistics about COVID-19, we also add some additional features such as a rich information map, flight searching with confirmed cases, manually filtered latest news and tweets, etc. As the founders of this project, we share our first-hand experience and lessons which may inspire further research on developing similar information dashboard for other emergency disasters, and new ways of education during the pandemic.  
\section{AU COVID-19 Dashboard Development}
In this section, we introduce the project characteristics, impact, and overall development process in steps. 
\subsection{Project Characteristic}
The first COVID-19 confirmed case in Australia was identified on January 25, 2020, and the increase was relatively slow in the next month.
Since the beginning of March 2020, the number of new COVID-19 cases were increasing significantly in Australia.
To keep people informed and calm, we started to develop an information dashboard to report the latest COVID-19 status in Australia on March 14, 2020, with a small team of the first three authors (one product manager and two developers).
We officially released our dashboard on March 16 with basic functionalities and open-sourced our project on GitHub\footnote{\url{https://github.com/covid-19-au/covid-19-au.github.io}}. It has become one of the first COVID-19 informative dashboards in Australia.
With the promotion and recruitment, our team increased from 3 members to 38 active members\footnote{\url{https://covid-19-au.com/about-us}} and over 200 volunteers from 12 different nations, most of whom are undergraduate/postgraduate students in Australia.
According to participants' expertise, we separated them into different groups, including web development, data/information/news collection, and communication.
Note that this project is nonprofit, and all participants volunteer to contribute to it. 

After the development, based on the existing dashboards from other countries and analysing user feedback with feature requirements and behaviors, we kept adding features to the dashboard. The information dashboard contains functionalities as below:
\begin{itemize}
    \item The latest number of new/total cases, death, recovery, testing in each state and territory.
    \item Data visualisation of historical data trend.
    \item A map of case distribution in each city or suburb.
    \item Flight details search with confirmed cases.
    \item Basic information about COVID-19 including symptoms, prevention, and a list of hospitals doing coronavirus testing.
    \item Latest news from online newsgroup and tweets from the Department of Health in State Government.
\end{itemize}

\begin{figure}
    \centering
    \includegraphics[width=0.48\textwidth]{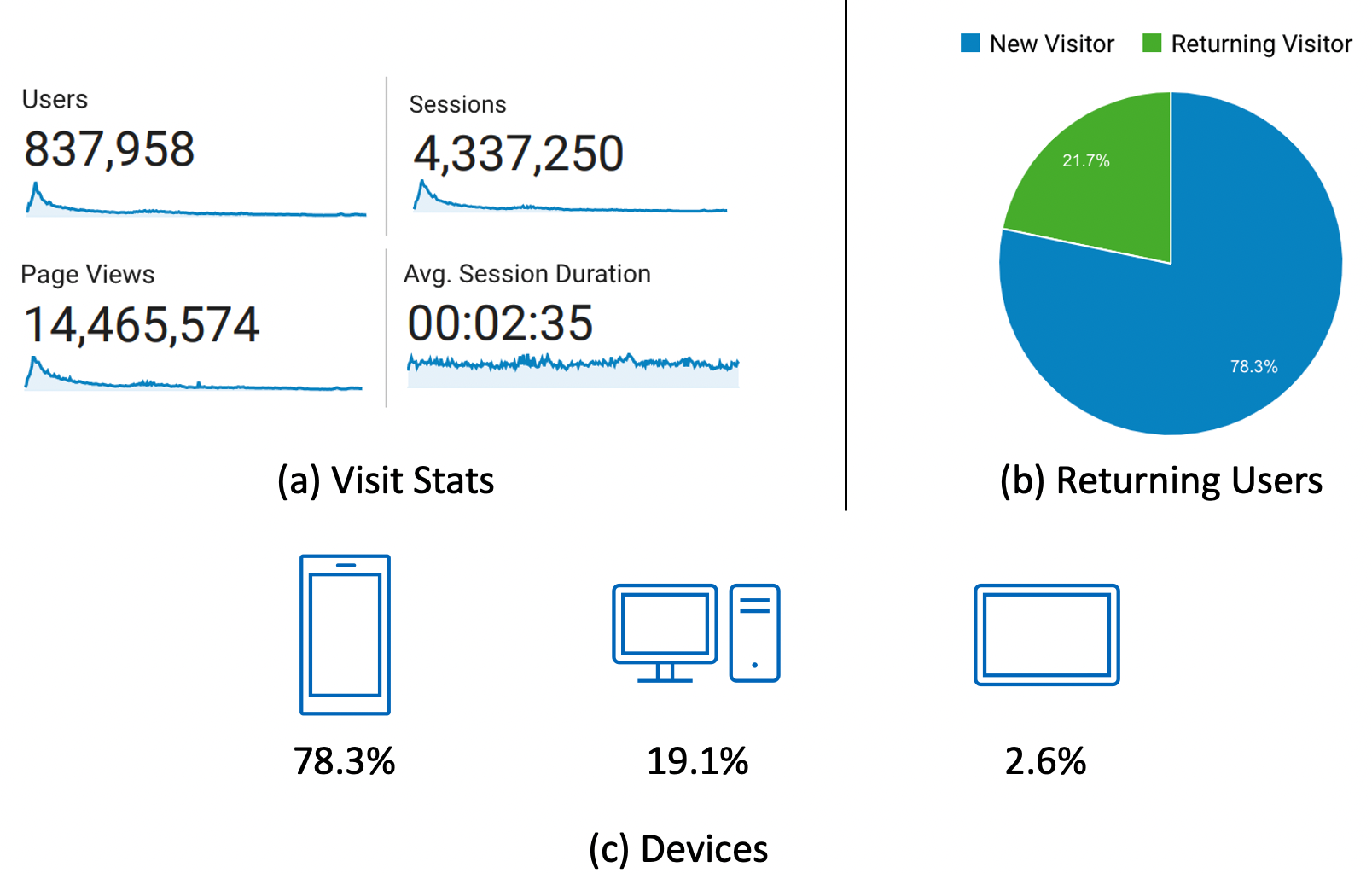}

    \caption{The visiting traffic of our dashboard}
    \label{fig:ga}
\end{figure} 
\subsection{Impact}
Our project is a social-good open-source project to keep people informed and calm during the pandemic.
Since the launch of this site, it has drawn a wide range of attention across Australia.
From March 17, 2020 to July 16, 2021, there have been 837,958 individual users with 14,465,574 page views, as seen in Fig~\ref{fig:ga}. 
There are 4,337,250 sessions, and users spend over 2.6 minutes on average in each session. 
Among all the visitors, 21.7\% of the users are returning users, whereas the rest of them are new visitors. 
Most of the users (78.3\%) use mobile devices to access the dashboard, whereas 19.1\% of them use Desktop. Only 2.6\% of the users use a tablet to visit the website.
Although most users are from Australia(over 90\%), there are still some of them from other 184 different countries and areas around the world.
From different channels (e.g., social media, email, online survey), we received much feedback from users aged from 10 to 80, around 81\% of the users feel excellent or very good about our website. The feedback includes appreciation letters, bug/error reports, and feature requests.
For example, we received an appreciation letter from a Canadian user who has an immunosuppressed daughter lives in Australia: 

\say{\textit{The information that your site has provided brings my family reassurance that my daughter is in safe hands; and for that we are truly grateful.} } 

Furthermore, there is also feedback that expressed their appreciation while requesting more features like: 

\say{\textit{Thanks for all of your hard work to keep people informed about the spread of covid-19. Seeing the data each day really helps put everything in perspective and is an effective way to reduce the fear and panic that are being spread. The only addition I could think of would be to include a new version of the ``Cases, Deaths and Recoveries" graph but allow it to be broken down by each state.}}

Seeing such feedback really motivated the team to keep delivering features to the product and having a positive impact on the users.

\subsection{An Overview of Development Process}
\label{sec:context}
\begin{figure}
    \centering
    \includegraphics[width=0.5\textwidth]{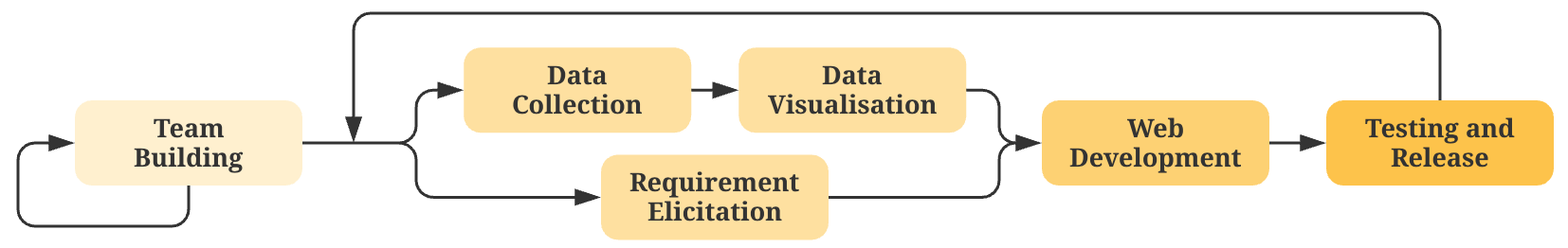}
    \caption{The development life cycle}
    \label{fig:process}

\end{figure}

In this section, we demonstrate the development process of the dashboard. Fig~\ref{fig:process} shows the development cycle of how we build and maintain the dashboard. Our development cycle starts with team building. As most of our team members are volunteer students, they would need to focus on their class works first and then help with the development. So we have to constantly find new members to take place for some old volunteers who need to deal with their assignments first. Then, we adopt the Agile programming cycle in order to deliver value and response to the requirements frequently~\cite{beck2001manifesto}. The development team meets virtually online to discuss which feature to develop in the next sprint, and we only pick the tasks that are actually needed. Moreover, inside each sprint, the developers report their daily scrum online, pick tasks from Kanban, and do extensive code reviews and tests. We discuss the details of each step in the following part.

\textbf{(Step-1) Team Building } To develop this dashboard, we constructed a team initialised with the first three authors of this paper as founders of this project and gradually expanded to over 200 volunteers from 12 different countries (e.g., Australia, China, Vietnam, et cetera.) with different expertise (e.g., IT, business, health). Most of them are enrolled students who did not know each other before this project, and all of them are working as volunteers. We split the team into three groups: the data team (data collection and update), the development team (build and test the dashboard), and the communication team (manage the social media account and communicate with end-users). Each team collaborates closely with each other and keeps producing value during the development process.

\textbf{(Step-2) Requirement Elicitation } Different from conventional software development, we did not have a complete list of requirements before the project development and did not have enough time for user requirement analysis before we started the development due to the rapid change of the COVID-19 pandemic. Therefore, we designed the initial version of the website with basic functions by the inspiration from other dashboards developed in the counties with earlier cases (e.g., China, Singapore) and our development experience. After building and releasing the Minimum Viable Product (MVP) version of the website, the communication team keeps collecting feature requests from public users' comments and feedback. They then summarised the requirements and discussed with the development team to make implement plans. The plans include the new feature development and the improvements on the existing ones. 

\textbf{(Step-3) Data Collection } Data collection is a really important step to the information dashboard. The data needs to be accurate and trustworthy. The data collection team collects the data from different resources (e.g., Government websites, online newsgroups, social media) using a semi-automatic way to make sure the information users gained from our dashboard is correct and trustworthy. The automatic crawler extracts the data from the Internet, and volunteers double-check the results, including resolving the contradictory information, identifying the reliable resources, and make sure the numbers are correct. In terms of data trustworthiness, we only use the data from country and state government official release and their official accounts on social media. And for the online newsgroups, we use the news from mainstream national online newsgroups and big newsgroups in the states of Australia. We have our data sources listed here\footnote{\url{https://bit.ly/2IEdFNj}}. 

\textbf{(Step-4) Data Visualization } We understand that data visualisation is not just about display raw data directly to the public. With different types of data we collected in the previous step, we add different types of data visualisations. For example, we use the heatmap to show the case density in the suburbs and states, the line chart to show the historical COVID-19 trend in Australia and the other countries. We also use other charts like pie charts, bar charts, and stacked charts to display data, i.e., the age and gender distribution for each state in Australia. 

\textbf{(Step-5) Web Development } We develop the website with React JavaScript, code is stored at Github for code management, and the website is hosted on AWS. The development team adopts the Agile programming model. The team members pick tasks from Kanban and chat virtually through Slack channels or Zoom meetings regularly. They also chat with the data team and communications team to understand the requirement.

\textbf{(Step-6) Testing \& Release } After the implementation of each new feature, we manually test our web applications on different devices to make sure the features work on all popular operating systems and different screen sizes. We also ask a small portion of our users to test the features and send feedback to us. After passing the tests, we release the new features immediately to meet the public needs on time.

\section{Lessons learned in dashboard development}
\label{sec:lesson}
During the rapid developments, we have made some mistakes as well as conducting some experiments with our users. We also collect and analyse the user activities to learn how they are using the COVID-19 dashboard to continuously improve the website. With our first-hand experience and theories in software development, we summarise some lessons that are highly related to the pandemic itself. Furthermore, based on our work duties(developers, researchers and instructors), we categorise the lessons into these three aspects, which should be helpful with people who work in the same areas respectively. 

\subsection{Lessons for Developers}
The software developers' productivity and wellbeing have declined due to the pandemic~\cite{ralph2020pandemic}. To help them with their development in future crises, we summarised four lessons for the developers based on the mistakes we have made and the experience we gained in the development process. The lessons are related to data collection and visualisation, entity naming, and prioritise development tasks.

\begin{figure}
    \centering
    \includegraphics[width=0.5\textwidth]{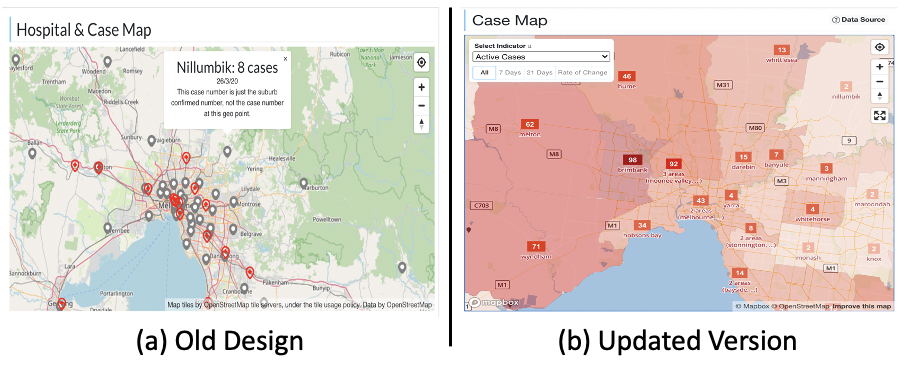}

    \caption{The update of case map }
    \label{fig:markermap}

\end{figure}

\begin{figure*}
    \centering
    \includegraphics[width=0.8\textwidth]{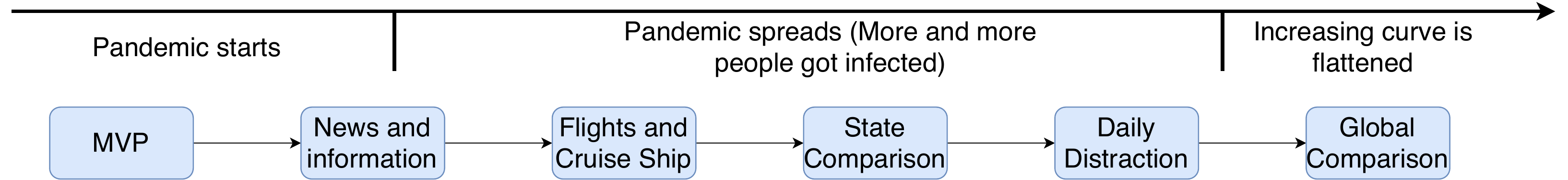}
    \caption{Temporal changes of our site}
    \label{fig:timeframe}
\end{figure*} 

\textbf{Lesson 1: Be careful about sensitive data}
Due to the sensitivity of the COVID-19 data, any errors in the data may incur social panic and spread fast among people~\cite{web:jhudataerror,web:jhudataerror2}. At the beginning stage of our dashboard, we have not built up a mechanism to efficiently update data in the dashboard. One day we accidentally put the death number wrongly by adding an extra zero after the number. 
Though we were aware of the error soon after the release,   we corrected the number and deployed the updated version as soon as possible, we still received a number of complaints (over 20) about the wrong number which resulting in users' confusion, panic, and disappointment of our data in the dashboard, which brought a negative influence to the dashboard.

To ensure the data accuracy, we then decided to adopt a human-centered semi-automatic mechanism to collect and verify sensitive data such as statistics and news, which kept both the data accuracy and maintenance update efficiency. In our approach, we first built the crawlers to fetch the latest stats from different trustworthy data sources. Then, we involved human into the loop (HITL)\cite{web:hitl} mechanism for having members from the data team to check the data accuracy and also resolving potential data conflicts. Finally, we published the data to the website for the public to view. Besides, we also simplified the deployment process to adapt quickly update so that if any unexpected errors happen, we can respond to them quickly. In such a way, we reduce the chances that errors might happen during the data collection process. 

Another example is when developing the cases map, we firstly use map markers to annotate COVID-19 virus stats in each city. As seen in (a) Fig~\ref{fig:markermap}, the red cross icons represent the geographic location of a COVID-19 clinic, and the grey markers represent the number of confirmed cases in that suburb. However, then we received feedback from users saying one of the markers was pointed right on his farm, which he was not happy with: 
\say{\textit{You have a case marked on our farm...}}. 
Furthermore, we also received feedback like the pointer was pointed in a national park which confused them if there was a virus case in the park. So we changed it to the (b) of Fig~\ref{fig:markermap}. We use the shades of red to represents confirmed cases in a certain area to avoid the ambiguous and misunderstanding.

\textbf{Lesson 2: Use domain-specific terms for a clear description}
In the development of such domain-specific app, it is crucial to choose professional terms rather than some random words which may cause ambiguity, confusion, and resulting in losing trust in the product. 
In our development experience of the COVID-19 dashboard, we once used the term ``\textit{Existing Case}'' to describe the confirmed cases that have not recovered yet. Though all of the developers in our team can understand this well, we received much feedback asking what does the \textit{Existing Case} refers to as they thought the \textit{Existing Case} is identical to \textit{Confirmed Case}, some even suspect that we were using fraud data.
We then realized that for this kind of cases, there is a more professional term as \textit{Active Case} for it. We quickly change the wording of it and add an explanation under the column to explain to the users who are still not clear about it. 
Furthermore, adopting correct medical terms which is both accurate and easy to understand is difficult, so we also further checked all the other words that we were using, compared them with the ones on the government Department of Health website, and consult our team members with medical backgrounds to make sure we are using the right form. 




\textbf{Lesson 3: keep users calm during the pandemic}
The COVID-19 virus not only has a physical impact on the patients but also brings heavy mental pressure, especially to people in lockdown or quarantine. 
There has been an increased number of reports of depression, anxiety, and distress across a number of medical staff and the public~\cite{mccall2020covid}. The social distancing, the isolated policies, and the fear of health and financial situations have negatively influenced people on their mental health. When developing the COVID-19 dashboard, we learned that the increasing number of death and confirmed numbers could cause people to feel anxious and worried. In order to avoid this, developers shall also consider keeping users calm when developing such dashboards.

Since knowing the facts about COVID-19 can help reduce stress~\cite{web:cdcStress}, in the COVID-19 dashboard, we develop the information page (seen in Fig~\ref{fig:overview}) that contains an instructional video of what the virus is, general information, and frequently asked questions about the COVID-19, and Australian state regulations. Furthermore, we also add a daily distractions section to the page, which includes videos, images, and news that are either fun and relaxing or motivated people during the time. 

\textbf{Lesson 4: Develop different features for different phases of the pandemic}
We develop the features that are only needed now~\cite{dande2014software}. Since the pandemic starts with no signs and it spreads around so quickly, there is no enough time for developers to develop all the features. Also, as our communication team keeps receiving feature requests from end-users, it is important to choose which features to develop under the current circumstances. In our practice, we chose the features that are needed the most based on different phases of the pandemic which keeps the continuous development, improves the efficiency of development, and increases developers’ motivation~\cite{dande2014software}. 

As can be seen in Fig~\ref{fig:timeframe}, we firstly developed a Minimum Viable Product(MVP) \cite{ries2009minimum}, based on the inspiration from other dashboards developed in other countries (e.g., JHU, DXY, etc). Then, we noticed that the Australian government started to release policies such as some necessary information about the virus and the regulations to the public. Moreover, some online newsgroups also had detailed news behind those numbers.  So we added the news and information features right after the MVP. As the COVID-19 pandemic was getting worse, we noticed that more than 60\% for the cases are related to overseas traveling\cite{web:sourceofIfection} so we updated the related flights and cruise ship information to the dashboard. As mentioned above, we also added the daily distraction part to keep people relax and calm during the pandemic. Finally, as the increasing curve was flattened, we noticed that the pandemic in other countries was still severe, so we added the global comparison to compare the pandemic in Australia with the rest of the countries to give the users a global view of the COVID-19 pandemic. 

In such a way, we managed to add new features in each new development iterations and keep delivering what is needed by the users at specific periods of the pandemic.

\begin{center}
 \begin{tcolorbox}[colback=black!5!white,colframe=black!75!black]
\textbf{Summary of lessons for developers}: Software developers play an important role in fighting COVID-19. 
When rapidly developing an information dashboard for future crises, they should be careful about the sensitive data to avoid society panic, use domain-specific terms to avoid confusion, keep an eye on the mental health of users as well as delivering features based on different phases of the crisis.
 \end{tcolorbox}
\end{center}

\subsection{Lessons for Academic Researchers}
Apart from the development lessons shared with developers, we also collect much information about users' interactions and feedback to our information dashboard.
In this section, we explore the data we collected to distill lessons that may benefit other researchers' study and improve the development.

\textbf{Lesson 5: Understand people's information needs about COVID-19}
During the development of the dashboard, we have collected user feedback and their overall usage activities through Google Analysis.
By analyzing this kind of data, our platform can help other researchers and Governments better understand people's COVID-19 information needs to which they can work to cater to.


Based on the data we collected from Google Analytics\footnote{\url{https://analytics.google.com/}} which helps collect end-user behaviour of the website, we learned that except the home page, state pages like Victoria (512,413 times), New South Wales (164,192 times) and Queensland (80,125 times) are the most visited pages, which are also the top three states that have the highest confirmed case number of COVID-19 in Australia. Note that the pandemic in Victoria is especially severe compared with the rest two, which also makes its page view significantly higher compared with the others. We also collect how users interact with page components. 
Table~\ref{tab:activities} shows the interaction frequency with components. 
We can see that the most popular component in our dashboard is the Case Map, as COVID-19 is spread by human and people are most concerned with the spatial data.

\begin{table}
    \footnotesize
    \center    

    \setlength{\tabcolsep}{2em}
    \begin{tabular}{c |c}
        \hline
        \textbf{Component (actions)} &\textbf{Portion} \\
        \hline
        Case Map (click, zoom, span, etc) & 71.7\%\\
        Case by State (choose state) & 10.5\%\\
        Navigation (click) & 8.2\%\\
        Global Comparison (select countries) & 7\%\\
        News (direct to original news) & 1.5\%\\
        Flight Search (search) & 1\%\\
        \hline
    \end{tabular}
    \caption{Users activities data collected from Google Analytics}
    \label{tab:activities}

\end{table}

\begin{figure}
    \centering
    \includegraphics[width=0.5\textwidth]{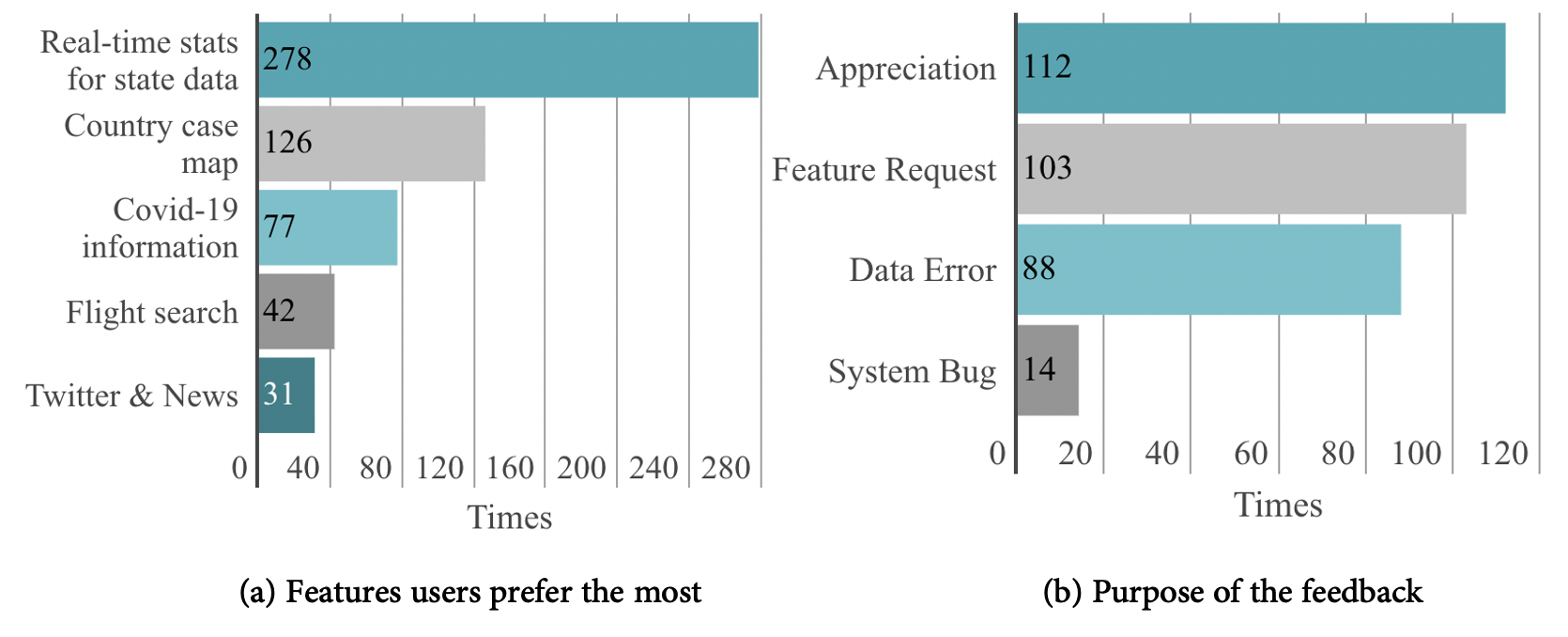}

    \caption{Results from the feedback}
    \label{fig:surveyResult}

\end{figure} 

For data from the user feedback, it is easier to figure out their information needs compares with the analytic tools, i.e. which part of the website do they like the most, what are the features they require. By September 2020, we have received a total of 315 feedback from users. As mentioned above, over 81\% of the participants feel excellent or very good about the website we build. 
Some other details as can be seen in Fig~\ref{fig:surveyResult}, of all the features in the dashboard, users prefer the real-time stats of the detailed state data the most. Most of the users come to express their appreciation to the dashboard and leave their feature requests or bug reports of the dashboard. We further cross-compare the data with their age and gender and find out some facts, i.e.  For users older than 40, they tend to raise bugs and data errors (37\% of users older than 40) whereas the younger ones are more likely to ask for new features (44\% of people younger than 40) such as adding the test number and the ICU number to the state table, introducing the log scale to the line charts, adding detailed age/gender distribution, etc. 

The data analysis from Google Analytics and users' feedback helps us to make a better adjustment to the dashboard in order to improve their user experience. For example, we noticed that a number of our users are using the Case Map to check the data with their mobile devices (over 80\% of users are mobile users), and we have also received some feedback expressed that it's hard to interact with the map on small screens. So, we added a full-screen option to the map for users to be able to better navigate on the map (as can be seen in Fig~\ref{fig:markermap} as well). We also prioritised the feature requests based on the number of them accordingly.

\begin{figure}
    \centering
    \includegraphics[width=0.4\textwidth]{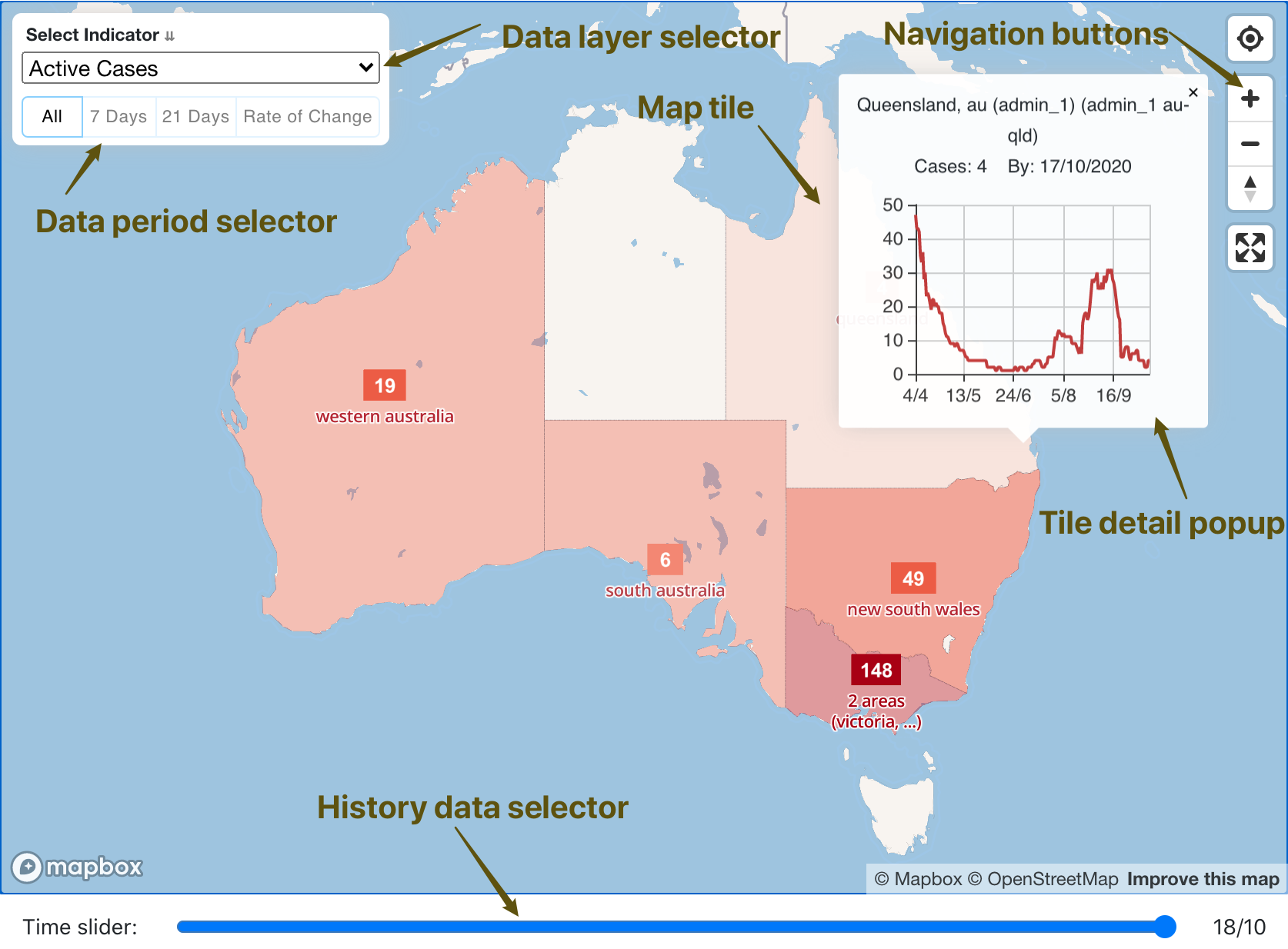}
    \caption{The interactive data map}
    \label{fig:mapExplained}

\end{figure}

\textbf{Lesson 6: Understand users' different interactions on mobile and desktop devices}
The interactive COVID-19 case map, as illustrated in Fig~\ref{fig:mapExplained}, has been considered to be one of the most critical parts of our dashboard according to our user feedback (as Case map in Table~\ref{tab:activities}). 
Epidemics of emerging infectious diseases such as COVID-19 are often observed as several clusters of disease at widespread locations \cite{thrusfield2005foot}, and its transmission is more likely to occur between individuals that are in close proximity \cite{pfeiffer2008spatial}. 
Therefore, spatial data such as confirmed case numbers in different areas are quickly disclosed and broadcast to the public. 

When people perceive such data, they naturally interact with it as it is easy to understand. This is one of the reasons for the case map component is popular.
However, due to the complexity of the spatial data and the characteristic of this pandemic, we hope to provide more insights to researchers on how users interact with the map especially in mobile platforms, which can help with their system design in future crises. 


Users can interact with the map by actions including zooming, panning, and clicking. We analyzed 779,018 records of our dashboard's map interaction logs from August 10 to September 9, 2020. These logs were further aggregated into 61,322 sessions.
Among these sessions, we found that more of them were generated by mobile users than desktop users. 43,503 sessions were carried out by mobile devices, which took 70.94\% of all the sessions. Moreover, there was a clear difference between the interaction patterns of the two user groups. 
Mobile users spent less time interacting with the map. A mobile session's average duration was 41.57 seconds, while it was 94.61 seconds for a desktop user. 
Correspondingly, mobile users generated fewer activities.
Their average number of activities per session was 11.57 times, which was 25\% less than desktop users.
The reason might be that it was hard for mobile users to carry out complicated gesture actions on the small screen than desktop users with a mouse or touchpad.

\begin{figure}
    \centering
    \includegraphics[width=0.4\textwidth]{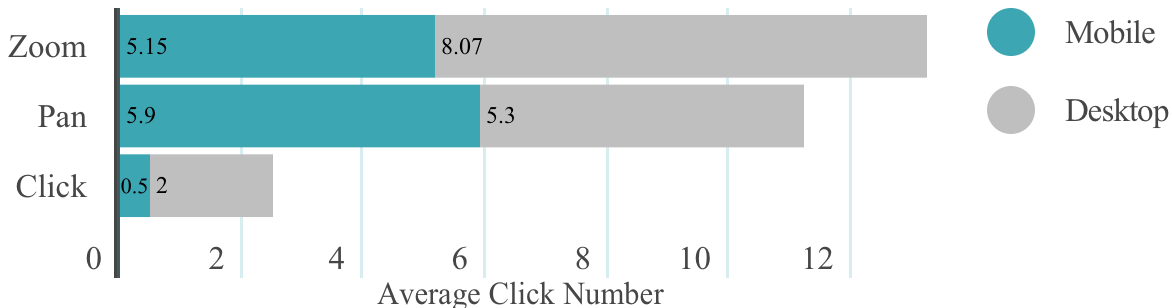}
    \caption{Session activities breakdown}
    \label{fig:activities}

\end{figure}

Navigation took up the majority of all the activities. As indicated in Fig~\ref{fig:activities}, mobile users and desktop users spent 95.59\% and 86.99\% of their activities respectively on zooming and panning. After users locate an area, they can click on it to see the detail. 
Mobile sessions had a 0.51 average click volume, which means that only 50\% of them click on any of the tiles. 
In comparison, desktop sessions had 2 average clicks. 
Most users were not enthusiastic about finding the tile detail, including history numbers and trends. 85.91\% of all the map sessions had not clicked any tile on the map. However, when calculating the sessions with at least one click recorded, we found that those sessions had 4.2 clicks on average. This indicates that those who want to explore the details tend to select multiple places inside one session. 

\textbf{Lesson 7: Understand users' interests in the map}

Apart from the interactions, researchers can further tell users' interest areas on the map by mining different interaction records.
The results may help the Government make the policies accordingly, understand one area's mental stress, etc. 

From the data we collected from the case map, we find that the hot area in the map is proportional to the number of active cases in the areas.
Refer to Table~\ref{tab:clicksOnArea}, of all the 8 Australian states showing on the map the most clicked one is Victoria, which is the state that also has 74\% of all the cases in Australia by October 2020. The Local Government Area (LGA) level data shows the same pattern. The most clicked tile is Wyndham Victoria, which has the most COVID-19 cases among all the other LGA areas. 
Out of the 14,168 sessions that recorded a click, 71.2\% of them clicked on a tile that is 50km away from their IP locations, which indicates apart from where they are based, the users are also interested in the situation outside the place they live in or work at. 
This reveals that users show more interest in the hot spot area with more case numbers regardless they are the residents or not.


\begin{table}[]
    \footnotesize
    \center    
    \setlength{\tabcolsep}{1em}
\begin{tabular}{c|c|c}
\hline
    \thead{\textbf{State}} &\thead {\textbf{Clicks}} &\thead {\textbf{Case Number}}\\
    \hline
    \multirow{1}{*}{Victoria} & 4,849   & 20,269\\
    \multirow{1}{*}{New South Wales } & 2,625   & 4,273\\
    \multirow{1}{*}{Queensland } & 1,395   & 1,161\\
    \multirow{1}{*}{Western Australia } & 823    & 692\\
    \multirow{1}{*}{Southern Australia} & 500    & 473\\
    \multirow{1}{*}{Northern Territory } & 361    & 33\\
    \multirow{1}{*}{Tasmania   } & 191    & 230\\
    \multirow{1}{*}{Australian Capital Territory} & 27     & 113\\
    \hline
\end{tabular}

\caption{Clicks on state and territory tiles}
    \label{tab:clicksOnArea}

\end{table}

\begin{figure}
    \centering
    \includegraphics[width=0.5\textwidth]{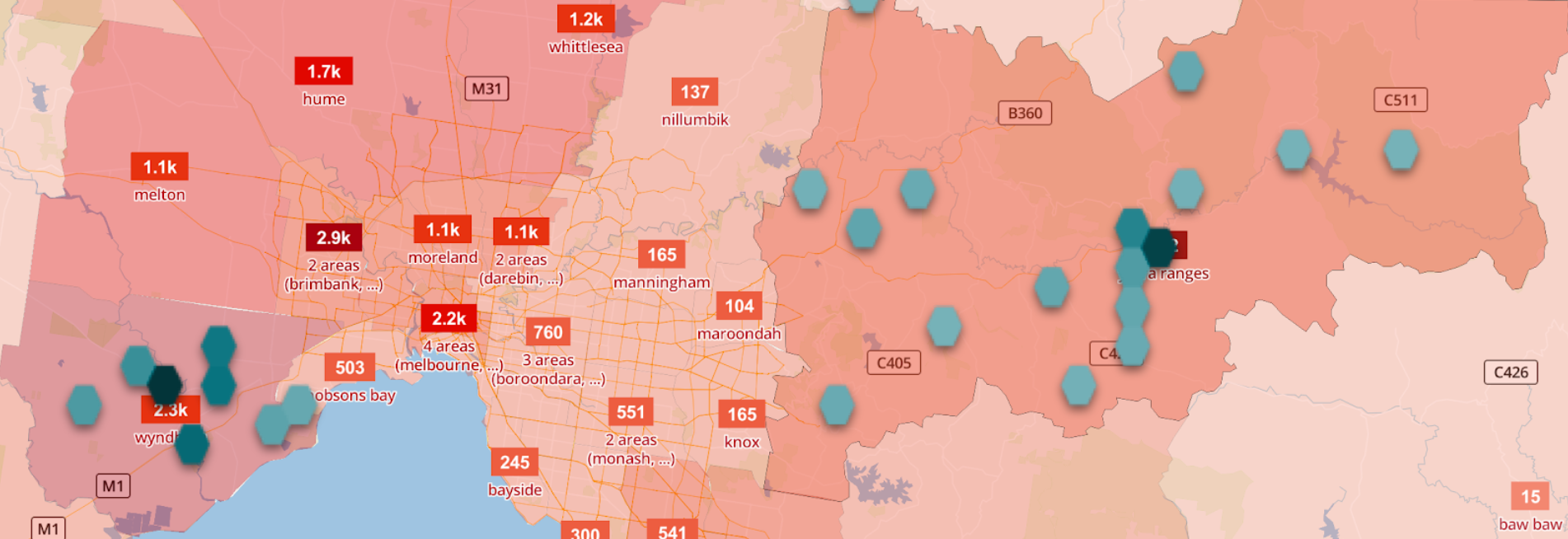}
    \caption{Clicks on Wyndham and Yarra Ranges aggregated (other clicks were hidden for clarity) }
    \label{fig:areasClicks}

\end{figure}

Another factor leading to the hot area is not related to the COVID-19 case number, but the size of that area. 
The Northern Territory State is of the least cases and least population among all the other states in Australia, but it attracted more clicks than Tasmania and Australian Capital Territory because it is the third-largest state in Australia in size. It is 15.6 times larger than Tasmania and 602 times larger than the Australian Capital Territory. A similar situation happens in clicks on LGA level data too. Take the second most clicked LGA area, Yarra Ranges Victoria as an example. Compare to the Wyndham (the most clicked LGA with the highest confirmed number), Yarra Ranges has 90\% confirmed cases less than Wyndham, but is more than 4 times larger than it. Fig~\ref{fig:areasClicks} shows the clicks on Wyndham and Yarra Ranges, the red layers represent the confirmed case numbers, whereas the green represents the click numbers. The darker the color is, the higher numbers it has. So the larger area gets a higher chance to be clicked.



The website also offers several data layer options for users to choose from (top left corner of Fig~\ref{fig:mapExplained}), i.e., active case, total case, total deaths, etc. However, only a few users appeared to have selected them. Excepting the default layer ``active cases" which was designed to be displayed for every session, the 4 most displayed layers (total cases, new cases, recovered number and death number) took 11\% of all sessions. A similar pattern was found in the history data periods filter and history data slider as well. Most sessions were not related to either of those functions.

\begin{center}
 \begin{tcolorbox}[colback=black!5!white,colframe=black!75!black]
\textbf{Summary of lessons for researchers}: By analyzing users' interactions and feedback to our information dashboard, we find that users always want to see more types of data and visualizations, they are interested in interact with case map the most, desktop users interact more compare with mobile device users, and they tend to focus on the area with high confirmed case numbers, etc.
The conclusion may help other researchers understand people during the pandemic and decision-makers in the government to design their policies.
 \end{tcolorbox}
\end{center}

\subsection{Lessons for Software Engineering Instructors}
Our dashboard is an open-source project initiated by Software Engineer (SE) instructors and run by volunteers, mainly university students. 
This section summarized the participants' experiences and listed several suggestions that may help SE instructors teach software engineering principles and concepts during the pandemic through project-based learning.

\textbf{Lesson 8: Students benefit from software engineering project-based learning in the pandemic} Although researches show that university students experienced heterogeneous disruptions during the pandemic \cite{aucejo2020impact, marelli2020impact, cao2020psychological}, our volunteers unanimously expressed that participating in the project is a beneficial experience. 

One of the benefits is that participating in such projects prepares them for their future career. Many participants revealed the gap between the industry requirement and skill they acquired from tertiary education \cite{radermacher2014investigating}. By participating in the project, they get a chance to apply their SE principles and knowledge to a real-world project. Our project is authentic, larger in scale, and more agile in pace compared to the capstone projects they were assigned from regular university courses. The site's development can be traced through its GitHub repository\footnote{\url{https://github.com/covid-19-au/covid-19-au.github.io}}. There were 29 contributors submitted 604 pull requests with 2921 commits in 206 days, which is approximately 14 commits per day. The two peaks of the commit volumes are aligned with the two waves of outbreaks in Victoria Australia, where the majority of the developers based. The commits were mainly submitted during the night, 62\% of the commits were submitted between 20:00 and 00:00 (GMT+11), as most of the major updates are deployed at night. The manual website updates last from morning to night, volunteers work in pair on shifts to submit and review data. The communication team runs multiple social media accounts, posts live data and write blogs to keep the public informed.

\begin{figure}
    \centering
    \includegraphics[width=0.45\textwidth]{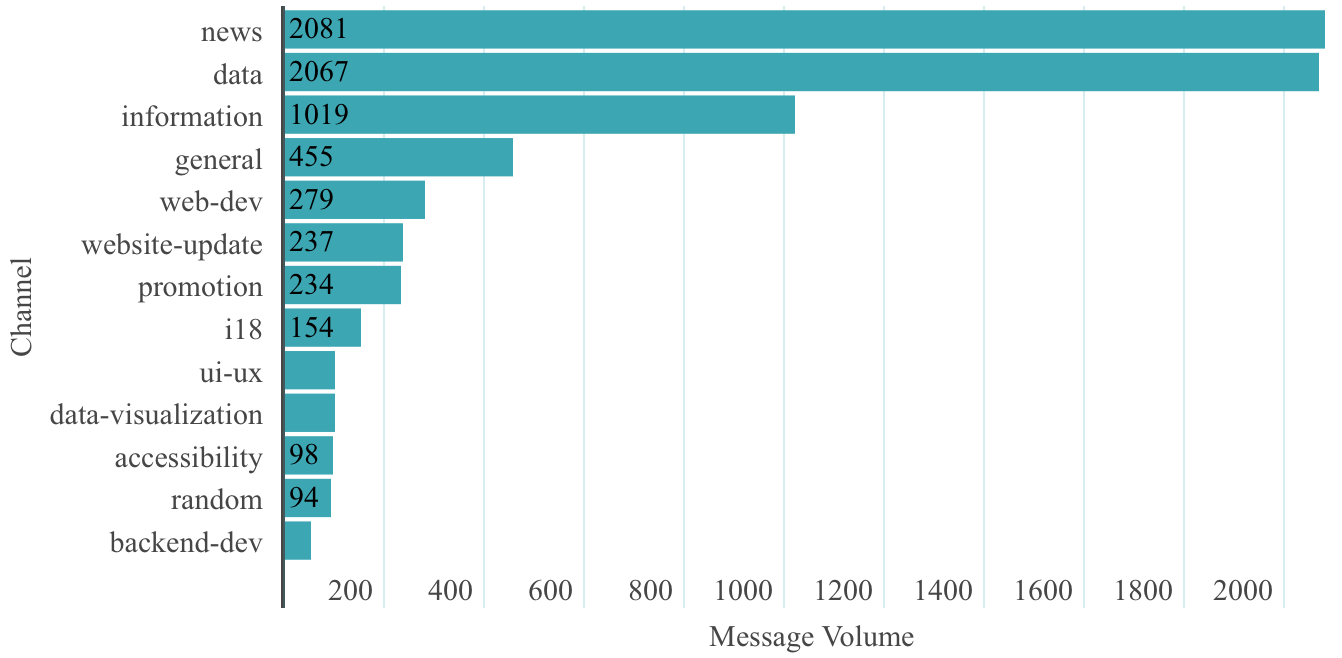}
    \caption{Slack message volume by channels}
    \label{fig:slackChannel}
\end{figure}

Being able to work with a big community closely is another benefit. It helps the students cope with the social distancing and lockdown, by providing a sense of belonging and equipping them with useful soft skills, especially the communication and cooperation skills in the large development team. In the project, over 200 members communicate closely to plan, implement, test, and maintain the dashboard to keep it abreast of the pandemic via slack across 13 different channels as per Fig~\ref{fig:slackChannel}. 
The discussions are mainly work-related. We applied the Latent Dirichlet Allocation(LDA)~\cite{blei2003latent} algorithm to all the chatting records and discovered the most frequently discussed topics are the data collection (includes news and case numbers), web page design, page testing, and requirements discussions. 

Therefore, instructors can use such a project-based learning approach to teach software engineering students under the epidemic allows students to collaborate within a large development group, and better understand the software development life cycle. 

\textbf{Lesson 9: Support students through the project process in the pandemic}

The COVID-19 pandemic has created many unprecedented challenges for students, and they need support from the instructors to better continue their study. Based on our experience during this open-source project, which is dominated by university students, we believe that instructors can provide assistance in the following areas.

Online open source project provides an immersive learning opportunity for the students, which is an effective pedagogy to teach Software Engineer principles and concepts. In our project, the instructor plays a role as the project manager. After completing the initial organization and helping the groups break the ice, the instructor guided the three student teams to independently arrange their rosters, schedule regular meetings, manage tasks and approach users. As a result, all the teams accomplished their works successfully and kept bringing values to the dashboard website. In a survey we conducted with another research group, students stated that both their technical and non-technical skills grew after participating in this project. The most significant growth from these two categories are front-end development (34\%) and working online skills (31\%). A number of the participants also mentions other skills like UI/UX (32\%), promotion (30\%), and time management(24\%).

\begin{figure}
    \centering
    \includegraphics[width=0.35\textwidth]{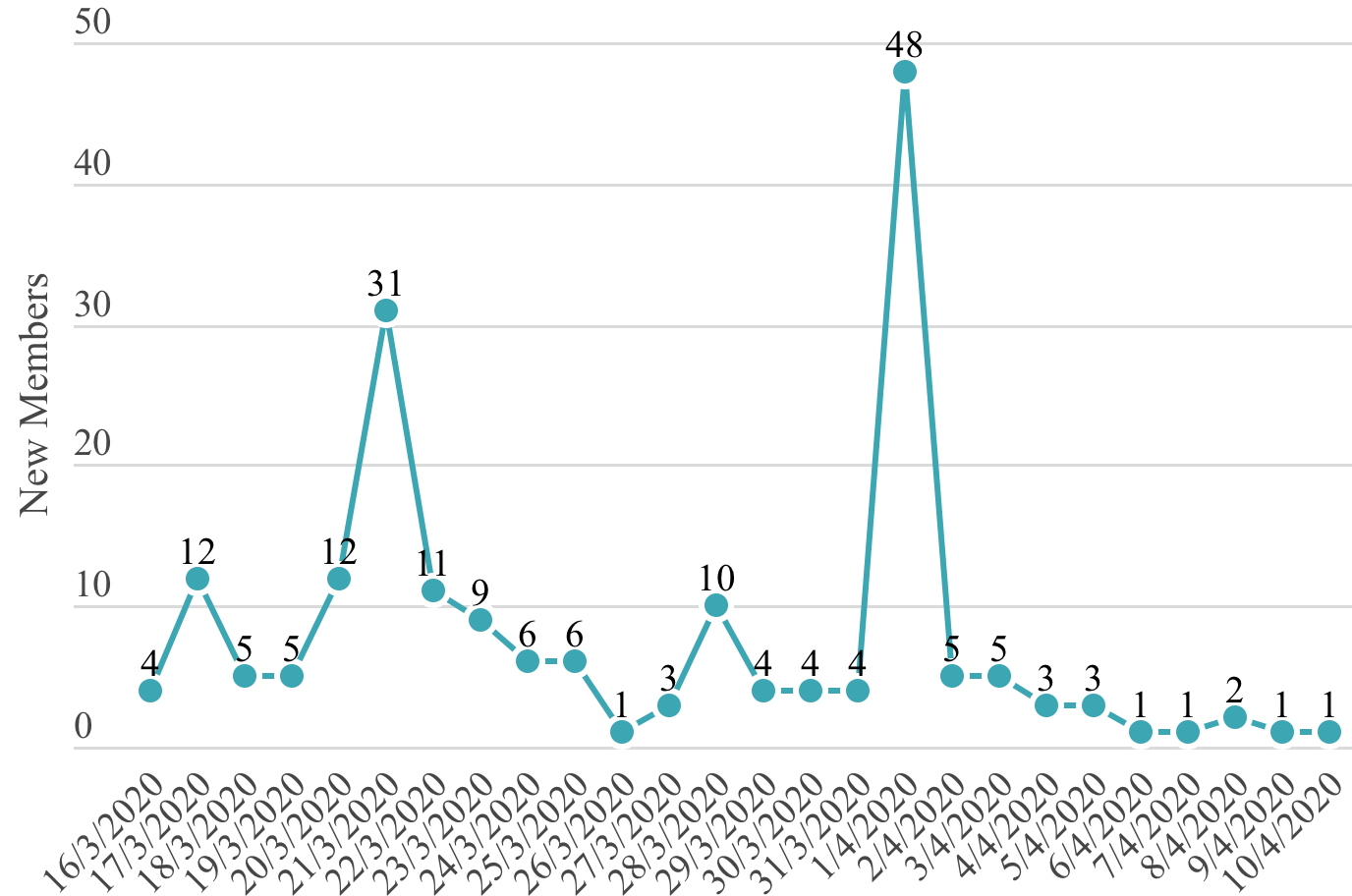}
    \caption{New recruitment overtime}
    \label{fig:slackOverview}
\end{figure}

Instructors' promotion can also help with the team building. Based our slack member analysis (Fig~\ref{fig:slackOverview}), we find that every time the volunteer team is enlarged in size is because of the instructors' help includes posts on the university website, Information Technology deportment newsletter, and class promotion. Many of the students revealed that they would not know they could participate into the project without such promotions, and they are grateful for this.

The sense of contributing to a great cause is the students' largest motivation, according to the student volunteers. Some explained that the project relieved them from powerless for not being able to do much during the pandemic apart from passively quarantined at home. However, the rapidly developing pandemic keeps causing pressure to the participants who are closely following it, which requires awareness from the instructors. We applied sentiment analysis to the Slack messages to see the atmosphere in our project and our instructor's strategy to keep members motivated while not overwhelming. An AFINN word list \cite{nielsen2011new} test shows the instructor has a 28\% higher sentiment level than the ordinary members. 
Of all the Emojis in all the chatting history, the instructor alone sent 16\% of them. This shows the instructor's attempt to use a positive attitude to motivate the group. Such efforts provide a friendly and encouraging atmosphere, which reduces the stress for students. Furthermore, we observed that the instructor is the most responsive to all the members' questions and achievements. This prevents the question owner's anxiety from building up while making people feel appreciated. 


\begin{center}
 \begin{tcolorbox}[colback=black!5!white,colframe=black!75!black]
\textbf{Summary of lessons for software engineering instructors}
The COVID-19 pandemic is changing the landscape of higher education, and instructors will have to adopt new ways to help students solve their challenges and create programs that are better suited for remote learning. Our experience has shown that organizing an open-source social-good software projects like ours is an good way for instructors to help students mitigate disruptive effects during the COVID-19 pandemic. By creating a positive environment and providing prompt support and feedback, instructors can help students gain experience and improve their skills, thus enhancing their motivation during development. 
\end{tcolorbox}
\end{center}

\section{Conclusion and Future Work}
\label{sec:conclusion}

In this project, we essentially followed an iterative process of understanding the users' values as the project emerged. We share our experiences on the development of COVID-19 Information Dashboard. Most importantly, we summarize 9 lessons for developers, researchers, and software engineer instructors, which may help them with their works and further benefit the society. Many of the lesson descriptions in this paper are about taking into account a particular human value during design. For example, lesson 3 concerns being sensitive to adverse physiological reactions in users because of the way data is presented. In an ideal world, the relevant values for an application would be uncovered upfront during requirements engineering. However, this is difficult in practice and a more iterative approach is required, as followed in this paper.Research shows that software engineers do not naturally think about human values during development~\cite{hussain2020human} and so the work here presents a case study from which useful insights may be drawn on how to properly understand and design for human values as the three roles a software engineer can be.

In the future, we are going to further explore the project in two directions.
On the one hand, we will further improve our site following the requirements from the public users and sharing more valuable data and experience with developers, researchers, and instructors.
On the other hand, we will try to abstract our existing project into a general template that can be adapted to other emergency disasters like bush fire, flood, etc.
We plan to release that template into an open-source platform for the community.

\bibliographystyle{ACM-Reference-Format}

\balance
\bibliography{reference}
\end{document}